\documentclass[11pt]{article}

\usepackage[english]{babel}
\usepackage{graphicx}
\usepackage{amssymb}

\usepackage{amsmath}
\usepackage{amscd}

\begin{document}
\newcommand{\be}{\begin{equation}}
\newcommand{\ee}{\end{equation}}
\newcommand{\bea}{\begin{eqnarray}}
\newcommand{\eea}{\end{eqnarray}}
\newcommand{\nn}{\nonumber}
\newcommand{\de}{\partial}
\hfill{\bf UB-ECM-PF-03/27}\par\hfill{\bf FIRENZE DFF-407/10/03}
\begin{center}
{\Large{\bf{}}}
\end{center}
\begin{center}
{\Large\bf\boldmath {Two tricritical lines from a Ginzburg-Landau
expansion: application to the LOFF phase}} \rm \vskip1pc {\large
R. Casalbuoni\footnote{On leave from Dipartimento di
 Fisica, Universit\`a di Firenze.}$^a$ and G. Tonini$^b$\\
\vspace{5mm} {\it{$^a$ Departament ECM, Facultat de Fisica,\\
Universita de Barcelona and Institut de F\'isica d'Altes
Energies,\\
Diagonal 647, E-08028 Barcelona, Spain}
 \\$^b$Dipartimento di
 Fisica, Universit\`a di Firenze,\\ I-50019 Firenze, Italia\\and\\
 I.N.F.N., Sezione di Firenze, I-50019 Firenze, Italia}\\
\vspace{5mm} email: casalbuoni@fi.infn.it, tonini@fi.infn.it}
 \end{center}
  \begin{abstract}

 \end{abstract}

 We study the behavior of the two plane waves configuration in the
 LOFF phase close to $T=0$. The study is performed by using
 a Landau-Ginzburg expansion up to the eighth order in the gap. The
 general study of the corresponding grand potential shows, under
 the assumption that the eighth term in the expansion is strictly
 positive, the existence of two tricritical lines. This allows to
 understand the existence of a second tricritical point for two antipodal
 plane waves in the LOFF phase and justifies why the transition becomes
 second order at zero temperature. The general
 analysis done in this paper can be applied to other cases.

\section{Introduction}

The condition for the  BCS  pairing is that the participating
fermions should have opposite momenta. This condition is realized
when the Fermi spheres of the two species are the same. However in
various circumstances this does not happen. For instance, in
ordinary matter, in the presence of a magnetic field the energies
of the spin up and spin down electrons are different and this
gives rise to a splitting of the two Fermi surfaces. In the case
of quarks a mass difference between the two pairing species leads
indeed to such a separation. The separation can also be produced
by the condition of weak equilibrium and/or charge neutrality. All
these different instances can be essentially described in terms of
a difference in chemical potential (real or effective),
$\delta\mu=(\mu_u-\mu_d)/2$, with $\mu_u$ and $\mu_d$ the chemical
potentials of the two fermions. As long as the separation is not
too large the BCS pairing still arises. However increasing the
separation leads to a first order phase transition to the normal
phase at a point known as the Clogston-Chandrasekhar
\cite{clogston} limit, given by
$\delta\mu=\delta\mu_1=\Delta_0/\sqrt{2}$, with $\Delta_0$ the BCS
gap.

Already many years ago Fulde, Ferrel \cite{FF} and Larkin,
Ovchinnikov \cite{LO} pointed out that close to this first order
transition a different pairing mechanism could arise\footnote{More
recently other possibilities have been considered, see
\cite{muther}}. In this case each of the fermions lie close to
their own Fermi surface and therefore the pair has non zero
momentum. This phase has been called both FFLO and LOFF in the
literature. Here we will adopt the name LOFF\footnote{For recent
reviews of this phase, see \cite{casalbuoni,bowers2}}. As a
consequence of the non zero momentum of the pair  translational
and rotational invariances are broken. It turns out that a
possibility is the formation of crystalline structures with the
vertices of the cell identified by the possible momenta of the
pair \cite{LO}. Correspondingly the condensate has a spatial
dependence of the type \be \Delta({\bf r})=\sum_{m=1}^P\Delta_m
e^{{2i{\bf q}}_m\cdot{\bf r}}.\ee

The case of a single plane wave has been studied both in \cite{FF}
and \cite{LO}. In \cite{LO} two other cases were considered: the
case of two antipodal plane waves and the one corresponding to a
cube. All these studies have been performed at zero temperature,
$T=0$. More recently this problem has become of interest in QCD
\cite{alford} just for the reasons mentioned above, that a
splitting in the Fermi energies can be produced by different quark
masses and/or by weak equilibrium and charge neutrality. In
\cite{bowers} 23 different crystalline structures have been
studied (still at $T=0$) with the conclusion that the most
favorable structure would correspond to the face-centered cube,
described by eight plane waves.

As far as the studies at $T\not=0$ are concerned, the phase
diagram in the plane $(\delta\mu,T)$ illustrating the  transition
from the BCS phase to the normal phase can be found in
\cite{sarma}. The main result is the existence of a tricritical
point, that is of a point where a first-order transition line and
a second order one meet. This is easily understood since at
$\delta\mu=0$ we have the usual BCS second order transition at
$T_c\approx 0.57\Delta_0$, whereas at $T=0$ we have a first order
transition at $\delta\mu=\Delta_0/\sqrt{2}$. The tricritical
point, $P_{tric}$ occurs at $\delta\mu_{tric}\approx 0.61\Delta_0$
and $T_{tric}\approx 0.32\Delta_0$ \cite{buzdin,combescot}. It
turns out that this tricritical point is the end point also for
any LOFF phase, in the sense that at that point also the vectors
${\bf q}_m$ vanish and we are back to a BCS-like situation. A
careful analysis of the LOFF phase has been done in
\cite{combescot}. The result is that it is very likely that the
favored phase close to the tricritical point is the one
corresponding to two antipodal vectors, that is to a condensate
\be \Delta({\bf r})=2\Delta\cos 2{\bf q}\cdot{\bf
r}.\label{due}\ee Furthermore the transition from this LOFF phase
to the normal one is first-order close to the tricritical point.
On the other hand, in \cite{LO} it has been shown that the
transition is second-order at $T=0$.

From the previous considerations we see that there are at least
two problems to understand, the first one is the possible change
of the favored crystalline structure from the tricritical point to
zero temperature and the second one is the change in the nature of
the transition in the case of the antipodal structure. A partial
answer to the first problem comes from two analysis made in the
two-dimensional cases \cite{shimahara,mora}. Essentially what
these authors find is that there is a series of transitions
starting from the antipodal case, close to the tricritical point,
between different crystalline structures. In particular, in ref.
\cite{mora} it is argued that the complexity of the crystalline
structure increases going toward $T=0$.

In the present paper we will be concerned with the second problem.
This has been studied in \cite{matsuo}  using the method of
quasi-classical Green's functions \cite{ashida}. The result is
that at very low temperature $T\approx 0.043 \Delta_0\approx 0.075
T_c$ there is a further tricritical point allowing the transition
from the first-order to the second order. In this paper we would
like to study this problem by using the more classical approach
based on the Ginzburg-Landau expansion around a second order phase
transition. The complication here is that we have to expand the
grand potential (or the free energy) up to the eight order in the
constant $\Delta$ of eq. (\ref{due}): \be\Omega=\alpha \Delta^2+
\frac{1}{4}\beta
\Delta^4+\frac{1}{6}\gamma\Delta^6+\frac{1}{8}\delta\Delta^8.
\label{potential}\ee In fact, whereas around $P_{tric}$ the
$\gamma$ coefficient is positive, and the expansion up to
$\Delta^6$ is enough to characterize the tricritical point,
$\gamma$ becomes negative going down in temperature
\cite{matsuo}. Therefore for reasons of stability of the theory a
further positive term in the expansion is necessary. Unfortunately
no calculation of the eighth order term exists so far, therefore
we will perform our study assuming $\delta$ positive up to zero
temperature and we will express our results parametrically in
$\delta$ itself.

Before studying the LOFF case we will consider the general phase
structure for a grand potential $\Omega$ as given in eq.
(\ref{potential}). Since we assume $\delta$ to be strictly
positive, $\Omega/\delta$ depends only on the ratio of the three
parameters $\alpha$, $\beta$ and $\gamma$ to $\delta$. We will
show that in this three-dimensional space there are two lines of
tricritical points. The point $P_{tric}$ lies on the line
$\alpha=\beta=0$, $\gamma>0$.  The other line is still in the
plane $\alpha=0$ but it lies on the $\gamma<0$ side and it is
given by the equation $9\beta\delta-2\gamma^2=0$.

As we have already noticed, from ref. \cite{matsuo} one can argue
that a second order line should exist starting from $T=0$ and
ending to another tricritical point. This point is located at low
temperature $T/\Delta_0\approx 0.043$. We will take advantage of
this fact by evaluating the parameters $\alpha$, $\beta$ and
$\gamma$ up to the second order in $T/\Delta_0$. Then one has  to
map the space $(\alpha,\beta,\gamma)$ on the parameter space of
the theory, that is $T$, $\delta\mu$ and the vector ${\bf q}$
characterizing the antipodal state to get the result. As already
anticipated we find a second tricritical point with a location
depending on the  value of $\delta$. We recover the results of
ref. \cite{matsuo} for very small values of $\delta$. In
particular the temperature and the $\delta\mu$ values
characterizing this point vary of about 25\% and 1.5\% in the
range $0\le\delta\le 1$. Therefore we confirm the results of the
analysis made in \cite{matsuo}, as long as the parameter $\delta$
is strictly positive.

We would like to point out  the  analysis we have made of the
critical points of the grand potential given in eq.
(\ref{potential}) is completely general and as such it can be
applied to all the situations where the eighth term in the
Ginzburg-Landau expansion plays a role.

In Section 2 we describe the Ginzburg-Landau expansion for the
LOFF case of two antipodal plane waves. Section 3 is dedicated to
the study of the critical points of the grand potential
(\ref{potential}). In Section 4 we apply the previous
considerations to the LOFF case. Conclusions are given in Section
5.

\section{The Ginzburg-Landau expansion for the case of two
   antipodal waves}

As discussed first in \cite{LO} and then in \cite{bowers} (see
also \cite{casalbuoni,bowers2}) one can derive the Ginzburg-Landau
expansion for the grand potential by first expanding the gap
equation in the gap parameter and then integrating over the
parameter itself. All the relevant equations are given in the
literature at zero temperature. The extension at finite
temperature is trivial and it is made by the introduction of the
Matsubara frequencies. As already said we will be concerned only
with the first three coefficients $\alpha$, $\beta$ and $\gamma$
which are the ones evaluated at $T=0$ in \cite{LO} and
\cite{bowers}, and we will assume the coefficient $\delta$ to be
strictly positive. Notice that we are  interested at the way in
which the first order transition line coming from the tricritical
point becomes second order close to zero temperature. Therefore it
will be enough to evaluate  the coefficients of the
Landau-Ginzburg expansion close  to $T=0$. For that we will
perform an expansion in powers of $T/\Delta_0$ where $\Delta_0$ is
the BCS gap, that is the gap evaluated at a value of $\delta\mu$
smaller than the Clogston-Chandrasekhar value,
$\delta\mu_1=\Delta_0/\sqrt{2}$.

The grand potential $\Omega$ is given in the GL approximation up
to the sixth order in the gap by
 \bea\Omega&=&-\frac 1 g
\Big( \sum_{k,n=1}^P\left[\Pi({\bf q_k},{\bf
q_n})\,-\,1\right]\Delta^*_k\Delta_n\delta_{{\bf q_k}-{\bf q_n}}
\cr&+&\frac 1 2 \sum_{k,\ell,m,n=1}^P J({\bf q_k},{\bf
q_\ell},{\bf q_m},{\bf
q_n})\Delta^*_k\Delta_\ell\Delta_m^*\Delta_n \delta_{{\bf
q_k}-{\bf q_\ell} +{\bf q_m}-{\bf q_n}}\cr &+&\frac 1
3\sum_{k,\ell,m,j,i,n=1}^P K({\bf q_k},{\bf q_\ell},{\bf q_m}
,{\bf q_j},{\bf q_i},{\bf
q_n})\Delta^*_k\Delta_\ell\Delta_m^*\Delta_j\Delta_i^*\Delta_n
\delta_{{\bf q_k}-{\bf q_\ell} +{\bf q_m}-{\bf q_j}+{\bf q_i}-{\bf
q_n}}\Big)\,,\label{gapGL} \eea where $P$ is the number of
independent plane waves in the condensate and \be\Pi({\bf
q_1},{\bf q_2})= \,+\,\frac{ig\rho}{2}\int \frac{d\bf\hat
w}{4\pi}\int_{-\delta}^{+\delta}d\xi\int_{-\infty}^{+\infty}\frac{dE}{2\pi}
\prod_{i=1}^2\, f_i(E,\delta\mu,\{{\bf q}\})\ ,\label{eq:143}\ee
\be J({\bf q_1},{\bf q_2},{\bf q_3},{\bf q_4})=
\,+\,\frac{ig\rho}2 \int \frac{d\bf\hat w}{4\pi}
\int_{-\delta}^{+\delta}
d\xi\int_{-\infty}^{+\infty}\frac{dE}{2\pi} \prod_{i=1}^4\,
f_i(E,\delta\mu,\{{\bf q}\})\ ,\label{gei}\ee\be K({\bf q_1},{\bf
q_2},{\bf q_3},{\bf q_4},{\bf q_5},{\bf q_6})= \,+\,\frac{ig\rho}2
\int \frac{d\bf\hat w}{4\pi}\int_{-\delta}^{+\delta}
d\xi\int_{-\infty}^{+\infty}\frac{dE}{2\pi}\prod_{i=1}^6
f_i(E,\delta\mu,\{{\bf q}\}),\label{kappa}\ee where, by putting
 ${\bf w}\,\equiv\,v_F\,\bf\hat w$,
 \be
 f_i(E,\delta\mu,\{{\bf q}\})=\frac{1}{E+i\epsilon\,
 \text{sign}\,E-\delta\mu+(-1)^i[\xi-2\sum_{k=1}^i(-1)^{k}
\bf w\cdot{\bf q_k}]}~.\ee Moreover the condition
\be\sum_{k=1}^M(-1)^{k}{\bf q_k}=0 \ee holds, with $M=2,4,6$
respectively for $\Pi$, $J$ and $K$. Furthermore $\rho$ is the
density of states at the Fermi surface given by $\rho=gp_F^2/\pi^2
v_F$, with $p_F$ Fermi momentum,  $v_F$ the Fermi velocity and $g$
the number of degrees of freedom. In the case of two-flavor QCD in
the color superconductive phase (2SC phase) one has $g=4$,
$p_F=\mu$ and $v_F=1$. If the vectors $\bf q$ define a crystalline
structure they belong to the orbits of the point group of the
crystal. In most of the cases considered in \cite{bowers} there is
a single orbit, and this is the case for two antipodal waves.
Therefore we assume that \be
\Delta_k=\Delta_k^*=\Delta~~~~~~(\text{for\ any\ } k)\,.\ee Then
we can rewrite (\ref{gapGL}) as follows (for two antipodal waves):
 \be \frac{\Omega}{\rho}=\alpha \Delta^2+\frac \beta 4 \Delta^4
 +\frac\gamma 6\Delta^6\,,\label{113}\ee
where \be \alpha=2\frac{1-\Pi(q)}{g\rho}, \ee
 \bea \beta&=&
-\frac 2{g\rho}\sum_{k,\ell,m,n=1}^P J({\bf q_k},{\bf q_\ell},{\bf
q_m},{\bf q_n})\delta_{{\bf q_k}-{\bf q_\ell}
+{\bf q_m}-{\bf q_n}} ,\label{beta1}\\
\gamma&=&-\frac 2{g\rho}\sum_{k,\ell,m,j,i,n=1}^P K({\bf q_k},{\bf
q_\ell},{\bf q_m} ,{\bf q_j},{\bf q_i},{\bf q_n})\delta_{{\bf
q_k}-{\bf q_\ell} +{\bf q_m}-{\bf q_j}+{\bf q_i}-{\bf
q_n}}.\label{gamma1} \eea These expressions, which are obtained in
\cite{bowers} for $T=0$, can be easily extended to finite
temperature by transforming the integration over the energy into a
sum over the Matsubara frequencies according to \be \int dE\to2\pi
i T\sum_{n=-\infty}^{+\infty}, ~~~E\to i\omega_n=(2n+1)\pi T.\ee
The way of evaluating these integrals at $T=0$ is explained in
\cite{bowers}. One can follow the same method and perform the sum
over the Matsubara frequencies using the Polygamma functions. One
can get the result in a finite form, but for our purposes is more
useful a low temperature expansion (in the ratio $T/\Delta_0$). In
fact, we will be interested in analyzing the line of second order
transitions starting from $T=0$. As anticipated in the
Introduction we expect this line to become first order at a
temperature of about $0.04\Delta_0$ \cite{matsuo}, therefore an
expansion up to the order $(T/\Delta_0)^2$ will be sufficient.

Notice that for any crystalline structure, the coefficient
$\alpha$ depends only on a single $\bf q$ vector. Therefore it has
a universal structure. In particular the second order points are
determined by the equation $\alpha=0$, and the optimal choice of
the vector $\bf q$ along a second order line is determined by the
condition \be\frac{\de\alpha}{\de{\bf q}}\Big|_{minimum}=0.\ee
This condition determines $|\bf q|$ leaving the direction of the
vector arbitrary. In particular, at $T=0$, one gets \be
qv_F\approx 1.2 \delta\mu_2,~~~\delta\mu_2\approx
0.754\Delta_0,\ee where $\delta\mu_2$ is the second order
transition point and $\Delta_0$ is the BCS gap. From this it
follows that all the vectors appearing in the quantities $J$ and
$K$ are of the same length, therefore due to the condition of
momentum conservation, only a few configurations may appear. In
the actual case of two antipodal plane waves and for the $J$
integral one has (with ${\bf q}_a=-{\bf q}_b$ and $|{\bf
q}_a|=|{\bf q}_b|=q$) \cite{LO} \be J_0=J({\bf q_a},{\bf q_a},{\bf
q_a},{\bf q_a}),~~~~J_1=J({\bf q_a},{\bf q_a},{\bf q_b},{\bf
q_b}). \ee It results \be \beta=-\frac{2}{g\rho}(2J_0+4J_1). \ee
In the case of $K$ we have three structures\bea &K_0=K({\bf
q_a},{\bf q_a},{\bf q_a},{\bf q_a},{\bf q_a},{\bf q_a}),
~~~~~K_1=K({\bf q_a},{\bf q_a},{\bf q_a},{\bf q_a},{\bf q_b}, {\bf
q_b}),&\nn\\&K_2=K({\bf q_a},{\bf q_a},{\bf q_b},{\bf q_b}, {\bf
q_b},{\bf q_b}),& \eea with \be \gamma=-\frac{2}{g\rho}(2 K_0+12
K_1+6 K_2). \ee We are now in the position of evaluating all the
quantities of interest at the order $(T/\Delta_0)^2$. About this
point, one can notice that all our expressions are even in $T$ and
therefore we are really neglecting terms of order
$(T/\Delta_0)^4$.

Introducing the re-scaled variables: \be T\to
\frac{T}{\Delta_0},~~~Q=qv_F\to
\frac{Q}{\Delta_0},~~~\delta\mu\to\frac{\delta\mu}{\Delta_0},\ee
 we get  the
following results \be \alpha(\delta\mu,Q,T)=-1+\frac{\delta\mu}{2
Q}\ln{\left|\frac{Q+\delta\mu}{Q-\delta\mu}\right|}-\frac{1}{2}
\ln{\frac{1}{4(Q^2-\delta\mu^2)}}-\frac{\pi^2
T^2}{3(\delta\mu^2-Q^2)},\label{a} \ee

\be J_0(\delta\mu,Q,T)=\frac{g\rho}{8}\frac{1}{ \delta\mu^2 -
Q^2}\left(1 +\frac{ \pi^2 (3 \delta\mu^2+Q^2)}{3 (\delta\mu^2-
Q^2)^2} T^2\right), \ee

\be J_1(\delta\mu, Q,T)=\frac{g\rho}{16\delta\mu}\left[\frac{1}{
Q}\ln\left|\frac{\delta\mu+Q}{\delta\mu-Q}\right|
+\frac{\pi^2}{3\delta\mu}\left(\frac{2(2\delta\mu^2-Q^2)}{
(\delta\mu^2-Q^2)^2}-\frac{1}{\delta\mu
Q}\ln\left|\frac{\delta\mu+Q}{\delta\mu-Q}\right|\right)T^2\right],
\ee

\be K_0(\delta\mu, Q,T)=\frac{g\rho}{64}\frac 1{(\delta\mu^2-
Q^2)^3}\left(3 \delta\mu^2+Q^2+\frac{ 2\pi^2(5 \delta\mu^4+10
\delta\mu^2Q^2+ Q^4)}{(\delta\mu^2- Q^2)^2} T^2\right), \ee

\bea &&K_1(\delta\mu, Q,T)=K_2(\delta\mu,
Q,T)=\frac{g\rho}{64\delta\mu^2}
\left[\frac{2\delta\mu^2-Q^2}{(\delta\mu^2-Q^2)^2}-
\frac{1}{2\delta\mu Q}\ln\left|\frac{\delta\mu+Q}{\delta\mu-Q}\right|\right.\nn\\
&&\left.+\left(\frac{2\pi^2(12\delta\mu^6-14\delta\mu^4Q^2+11\delta\mu^2Q^4-3Q^6)}
{3\delta\mu^2(\delta\mu^2-Q^2)^4}-\frac{ \pi^2}{\delta\mu^3
Q}\ln\left|\frac{\delta\mu+Q}{\delta\mu-Q}\right|\right)T^2\right].
\eea

We have now the expansion of the grand potential up to terms of
order $\Delta^6$. However for the following discussion one would
need also the next term in the expansion. Since this is not
available we will assume it to be strictly positive and we will
discuss the results as functions of the coefficient $\delta$ of
this term. Therefore, our grand potential will be \be
\frac{\Omega}{\rho}=\alpha \Delta^2+\frac \beta 4 \Delta^4
 +\frac\gamma 6\Delta^6+\frac\delta 8\Delta^8,\ee with $\alpha$,
 $\beta$ and $\gamma$ from the calculations of this Section and
 $\delta$ assumed strictly positive.

\section{The general phase diagram}

\begin{figure}[hbt] \centering
\includegraphics[height=5.8cm]{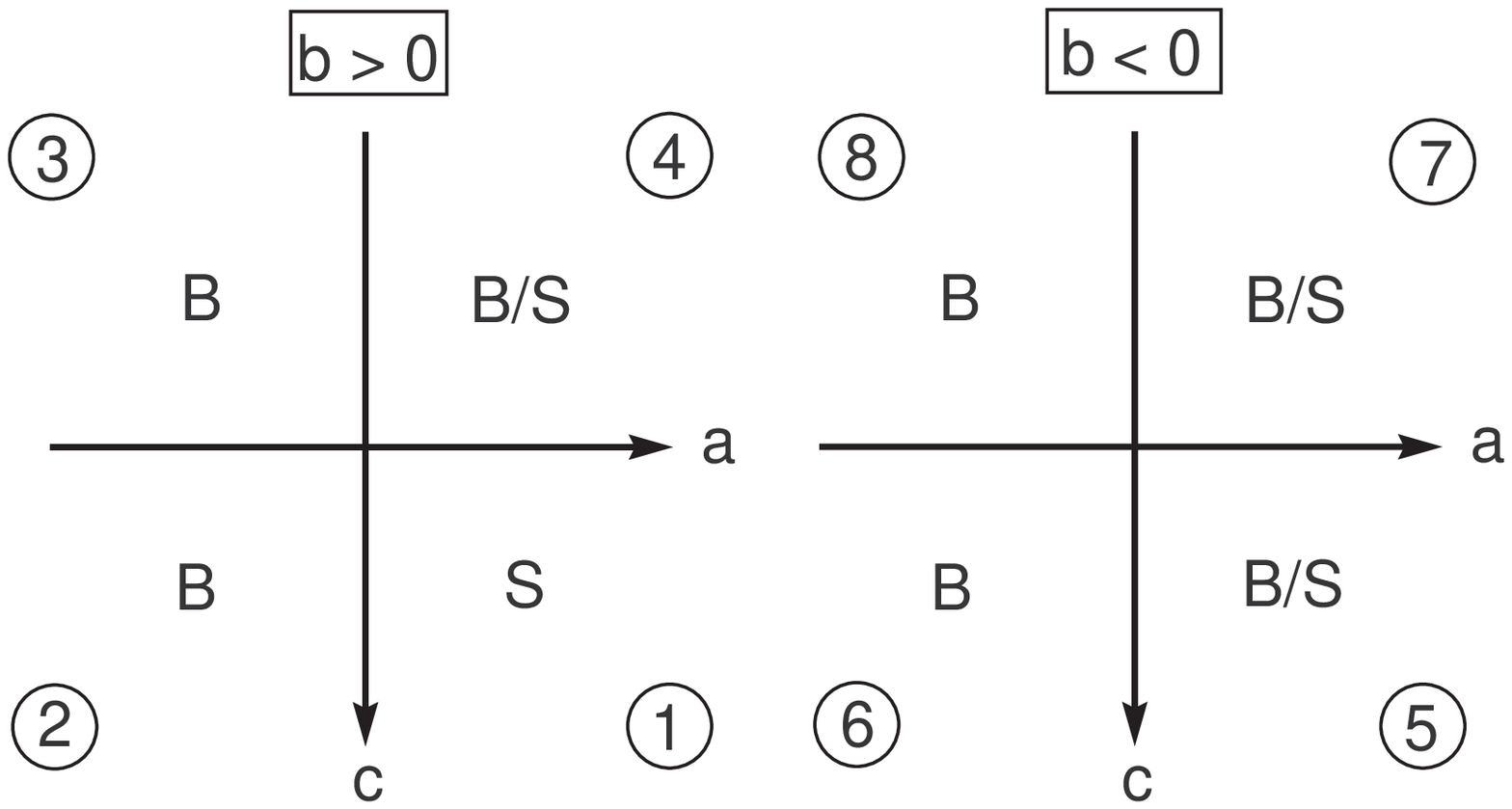} \caption{\it
The octants of the space $(a,b,c)$ are illustrated in the two
planes corresponding to $b>0$ (left panel) and $b<0$ (right
panel). Also the phases in the various regions are indicated. $S$
is the symmetric phase, whereas $B$ is the broken one.}
\label{ottante}
\end{figure}

To study the structure of the phase space let us define  a
dimensionless grand potential: \be
\bar\Omega=\frac{\Omega}{(\delta\Delta_0^6)\rho\Delta_0^2}=x\left(a+\frac
b4 x +\frac c6 x^2+\frac 1 8 x^3\right),\label{potential1}\ee
where \be
a=\frac{\alpha}{\delta\Delta_0^6},~~~b=\frac\beta{\delta\Delta_0^4},~~~
c=\frac{\gamma}{\delta\Delta_0^2},\ee  $\Delta_0$ is the BCS gap
and \be x=\left(\frac\Delta\Delta_0\right)^2.\ee The minima  of
the grand potential are given by the equation
\be\frac{\de\bar\Omega}{\de\Delta}=\frac{\sqrt{x}}
{\Delta_0}(2a+bx+cx^2+x^3).\label{potential2}\ee Notice that the
structure of the minima is not changed by dividing by the
parameter $\delta$ since it is assumed to be strictly positive.
Therefore our phase space is a three-dimensional one. We want to
determine the regions of this space corresponding to first and
second order transitions. Their intersection will fix the
tricritical lines. We start analyzing the roots of eq.
(\ref{potential2}) different from zero. The type of solutions with
their sign is given in Table \ref{table1}. Keeping in mind that we
have the constraint $x>0$ for not zero roots, we can easily
determine the kind of symmetry in each octant, and the result is
shown in Fig. \ref{ottante}.

\begin{table}[htb]
\begin{center}
\begin{tabular}{|c|c|c|c|c|c|}
\hline &&&&&\\  &  $a$ & $b$ &$c$ & Solutions & Phases\\&&&&&\\
\hline   1 & $+$ & $+$ & $+$ & 3 negative & symmetric\\
  &  &  &  & 1 negative, 2 complex & symmetric\\
\hline   2 & $-$ & $+$ & $+$ & 2 negative, 1 positive & broken\\
  &  &  &  & 2 complex, 1 positive & broken\\
\hline  3 & $-$ & $+$ & $-$ & 3 positive & broken\\
  &  &  &  & 2 complex, 1 positive & broken\\
\hline
  4 & $+$ & $+$ & $-$ & 2 positive, 1 negative & broken\\
  &  &  &  & 2 complex, 1 negative & symmetric\\
\hline
 5 & $+$ & $-$ & $+$ & 2 positive, 1 negative & broken\\
  &  &  &  & 2 complex, 1 negative & symmetric\\
\hline
  6 & $-$ & $-$ & $+$ & 2 negative, 1 positive & broken\\
  &  &  &  & 2 complex, 1 positive  & broken\\
\hline
  7 & $-$ & $-$ & $-$ & 2 negative, 1 positive & broken\\
  &  &  &  &  2 complex, 1 positive & broken\\
\hline
  8 & $+$ & $-$ & $-$ & 2 positive, 1 negative  & broken\\
  &  &  &  &  2 complex, 1 negative & symmetric\\
\hline \end{tabular}
\end{center}
\caption{The table shows the character of the solutions of the
cubic according to the sign of the coefficients. Also shown are
the phases corresponding to the various type of
solutions.}\label{table1}
\end{table}

In particular we see that for $a<0$ the symmetry is always broken.
This is confirmed also by looking at the second derivative of
$\bar\Omega$: \be \Delta_0^2\frac{\de^2\bar\Omega}{\de\Delta^2}=2a
+3bx+5c x^2+7 x^3.\ee The point $x=0$ for $a<0$ is always a
maximum, whereas for $a>0$ is a minimum, and we have to decide
which is the absolute minimum. If we evaluate the second
derivative at a root $x_r\not=0$ we find
\be\Delta_0^2\frac{\de^2\bar\Omega}{\de\Delta^2}\Big|_{x_r}=-12 a-
4bx_r -2c x_r^2,\ee which is negative in  the octant 1
($a,b,c>0$). Therefore in this octant the true minimum is at $x=0$
and we are in the symmetric phase as seen from the analysis in
Table \ref{table1}. To decide what happens in the octants 4, 5 and
8, the ones with mixed symmetry, it is enough to look at the first
order points. These are the points where, for $a\not=0$, we have a
change in the symmetry. Since $a\not=0$, the only possible minimum
in zero is the trivial one, and  these points are determined by a
change in the number of the real solutions of the cubic \be
\frac{\bar\Omega}x=a+\frac b4 x +\frac c6 x^2+\frac 1 8
x^3=0,\label{cubic}\ee as already evident from Table \ref{table1}.
Therefore the first order surface is made up by the points of the
regions 4,5 and 8 where the discriminant of the cubic equation
(\ref{cubic}) is zero. Before discussing this point let us notice
that the second order transitions are characterized by $a=0$, but
not all this plane is a second order surface. Regions that
correspond certainly to second order transitions is the part of
the plane $a=0$ separating region 1 from region 2. In the other
cases, that is 3-4, 5-6 and 7-8 one has first to discuss the
region of the first order transitions.

\begin{figure}[h] \centering
\includegraphics[height=5.8cm]{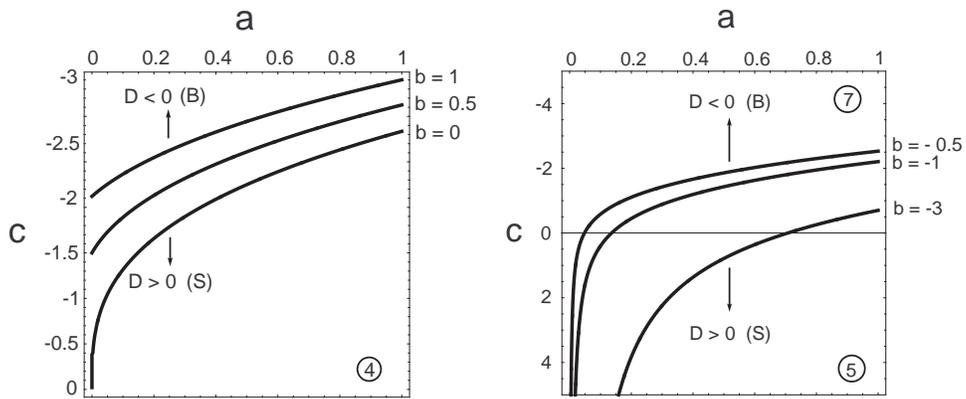} \caption{\it We show
the points where $D=0$ in the region 4, $a>0, b>0, c<0$ (left
panel) and in the regions 5 and 7 , $a>0, b<0$ (right panel). In
the part of the octants below these curves we have $D>0$,
therefore there are 2 complex and one negative root (see Table
\ref{table1}), and we are in the symmetric phase. By continuity
the region above the curves $D=0$ belongs to the broken phase.}
\label{separation}
\end{figure}

To discuss the solutions of (\ref{cubic}) we bring it to the
normal form  \be y^3+p y+q=0, \ee where \be y=\frac{4c}{9}+x\ee
and \be p=2b-\frac{16}{27}c^2,~~~
q=\frac{128}{729}c^3-\frac{8}{9}bc+8a.\ee The discriminant of the
cubic is then given by \be D=\frac{q^2}{4}+\frac{p^3}{27}\ee and
we have one real  and two complex solutions for $D>0$ and three
real solutions for $D<0$. Therefore for $D<0$ we are typically in
the broken phase. The situation is made clear in Figure
\ref{separation} where we show the curves $D=0$ in the regions 4,
5 and 7 for different values of $b$.

\begin{figure}[h] \centering
\includegraphics[height=12cm]{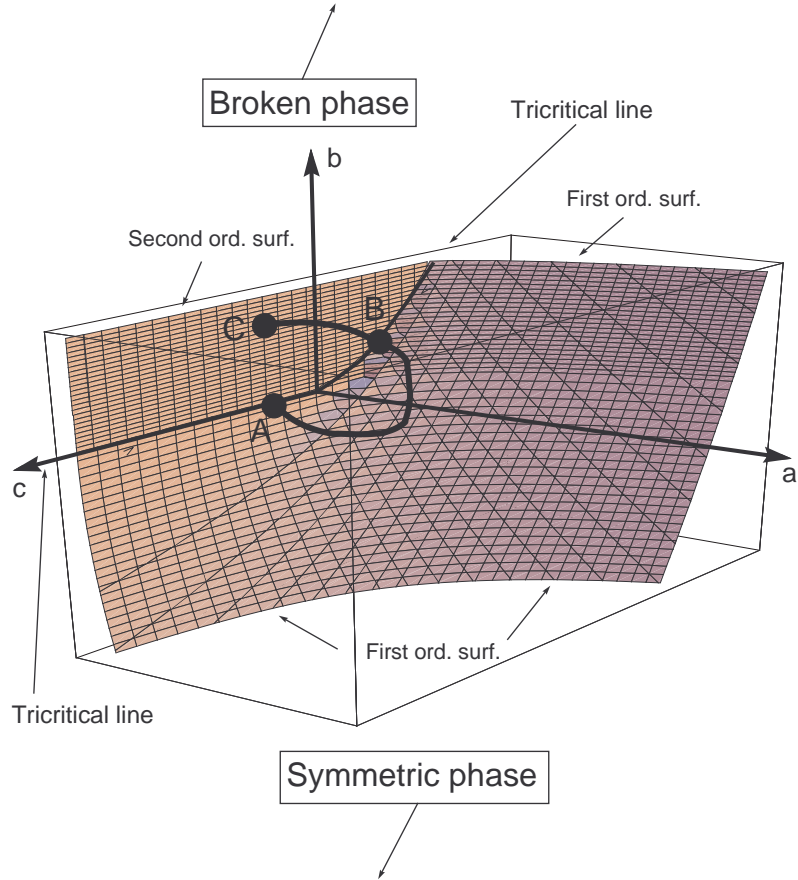} \caption{\it This three dimensional plot
shows the surface of first order points in the relevant regions 4,
5 and 7 and the surface of the second order points which lies in
the plane $a>0$ for $b$ and $c$ positive and in the part of the
plan $b>0$ and $c<0$ delimited by the surface of the first order
points. Also shown are the tricritical lines. In this figure we
show also roughly the path along the critical line starting at the
tricritical point (A) of the LOFF phase down to zero temperature,
point C. In the path the other tricritical point B is found. The
rest of the path between B and C  is along the second order
surface. The arrows in the boxes containing broken and symmetric
phases indicate that these two phases are respectively behind and
in front of the critical surfaces.} \label{3dim}
\end{figure}

We can now identify the tricritical lines noticing that they are
necessarily at the intersection of the plane $a=0$ with the
surface $D=0$ passing through the regions 4, 5 and 7. Evaluating
$D$ at $a=0$ we get: \be D|_{a=0}=\frac 2 {27} b^2\left(b-\frac 2
9 c^2\right).\ee The line $a=0$ and $b=0$ for $c>0$ is certainly a
tricritical line, since it belongs to both to  the first order and
the second order surfaces. As far as the line $b=2c^2/9$ it can be
critical only for $b>0$ and $c<0$ (region 4). In fact,  in this
case the first order lines touch the surface $a=0$, as shown in
Fig. \ref{separation}, whereas in the regions 5 and 7 this does
not happen (see again Fig. \ref{separation}). Summarizing the
tricritical lines are defined by\be \alpha=\beta=0,
~~~~\gamma>0\ee and \be 9b-2c^2=0,~~b>0,~~c<0.\label{42}\ee The
first and second order critical surfaces with their intersections
are shown in the three dimensional plot of Fig. \ref{3dim}.

\section{The second tricritical point for the two antipodal plane
waves}

\begin{figure}[h] \centering
\includegraphics[height=6cm]{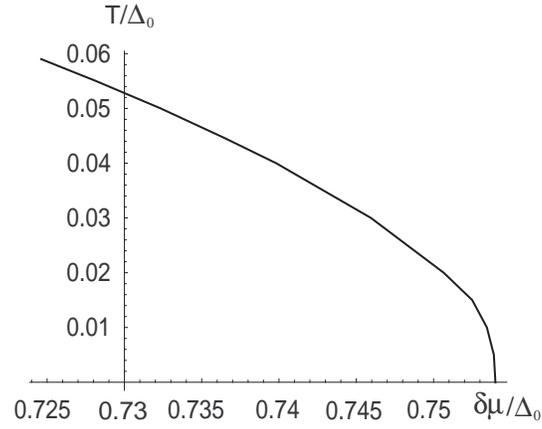} \caption{\it This figure
shows the second order transition line in the plane
$(\delta\mu/\Delta_0, T/\Delta_0)$.} \label{phasedi}
\end{figure}

\begin{figure}[h] \centering
\includegraphics[height=6cm]{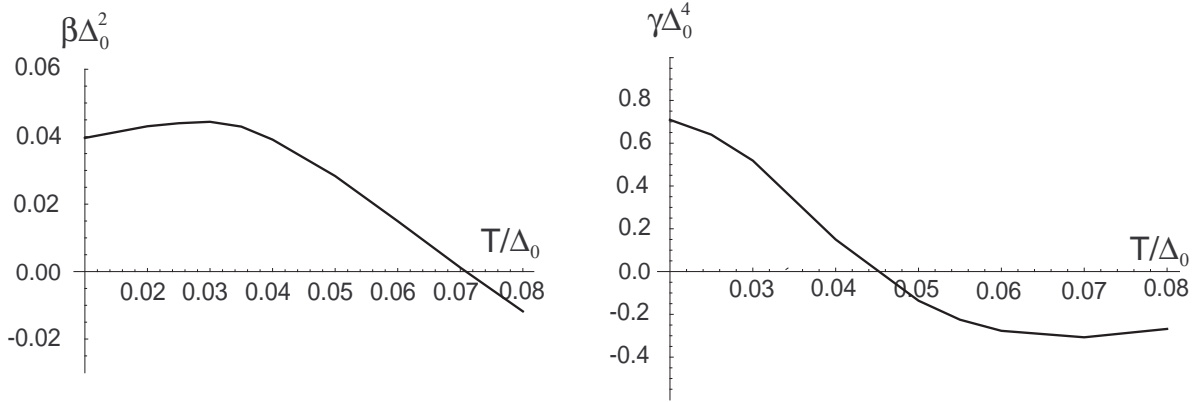} \caption{\it The dimensionless
quantities $\beta\Delta_0^2$ (left panel) and $\gamma\Delta_0^2$
(right panel) vs. $T/\Delta_0$ along the second order transition
line.} \label{betagamma}
\end{figure}

\begin{figure}[h] \centering
\includegraphics[height=12cm]{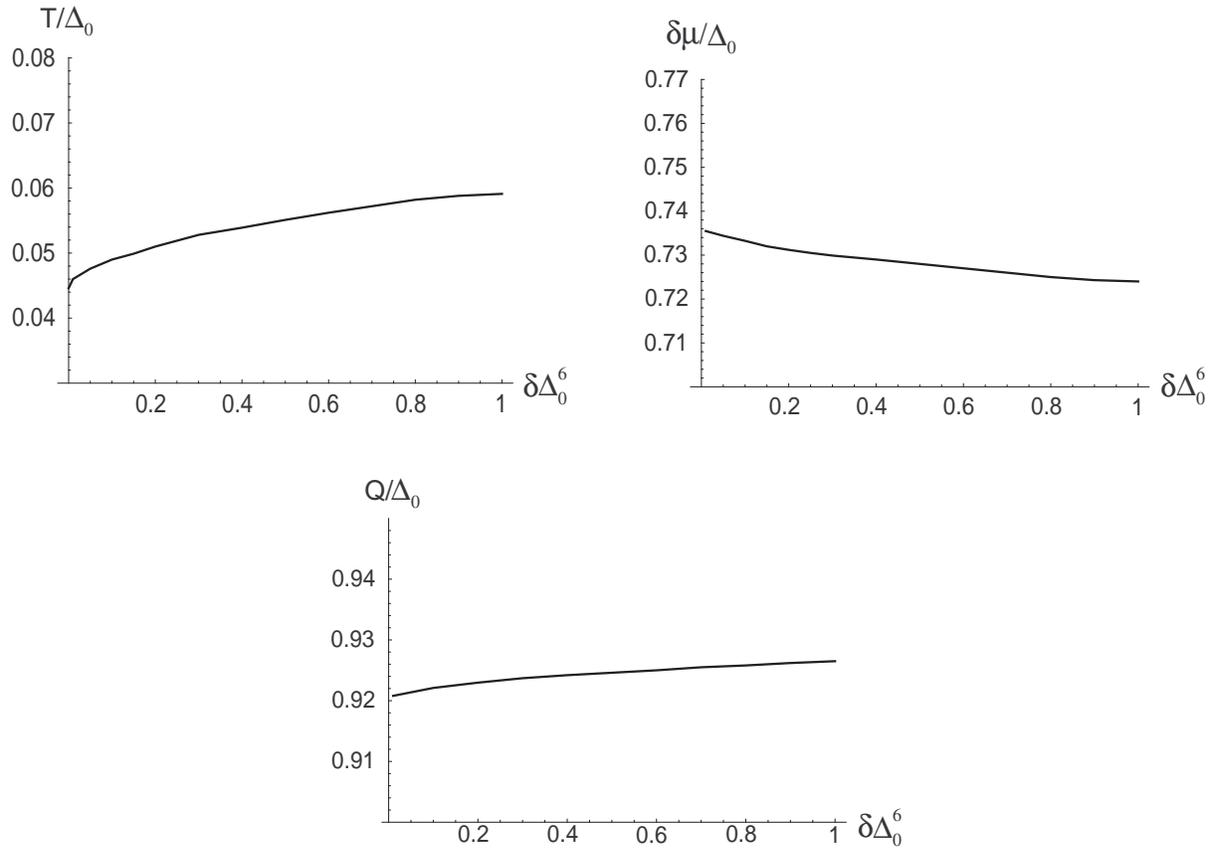} \caption{\it The figure shows the variation
of the position of the second tricritical point obtained by
solving eq. (\ref{secondtr}) in $T$ (left-upper panel) and
$\delta\mu$ (right-upper panel). Also shown is the optimal value
for the wave vector $Q$ at the tricritical point vs. $\delta$
(lower panel).} \label{delta}
\end{figure}

We have now all the elements to analyze what happens in the case
of two antipodal plane waves along the second order transition
line starting from $T=0$. These points, as we have discussed in
the previous Section are defined by the condition \be
\alpha(\delta\mu,T,Q)=0,\ee supplemented by the condition of
optimality for $Q$ which amounts to require the stationarity of
the grand potential $\Omega$ \be
\frac{\de\alpha(\delta\mu,T,Q)}{\de Q}=0.\ee We can solve these
two equations finding the values of $Q$ along the second order
transition line and a relation between $T$ and $\delta\mu$. Of
course, the result is the same as in the case of a single plane
wave, since as noticed before, the expression of $\alpha$ does not
depend on the number $P$ of plane waves. The transition line in
the plane $(\delta\mu,T)$ is given for small values of $T$ in Fig.
\ref{phasedi}. What is different now from the case of the single
plane wave is the behavior of the coefficients $\beta$ and
$\gamma$. In fact, looking at Fig. \ref{3dim}, for the single
plane wave one has a path going from $C$ ($T=0$) to $A$ (the
tricritical point) staying along the second order surface, whereas
in the actual case one moves from the region $\beta>0$ and
$\gamma>0$ to the region $\beta>0$ and $\gamma<0$ where the second
tricritical line (see eq. (\ref{42})) \be
9\beta\delta-2\gamma^2=0\label{secondtr}\ee is met. This is shown
in Fig. \ref{betagamma}  where the behavior of $\beta$ and
$\gamma$ along the second order transition line is shown as a
function of the temperature. In particular we see that for $
0.045\Delta_0\lesssim T\lesssim 0.071$, $\beta$ is positive and
$\gamma$ is negative. Therefore in this region another tricritical
point is found. The transition becomes first order and stays like
that till the first tricritical point $A$. The location of the
second tricritical point is given by equation (\ref{secondtr}) and
the results are illustrated in Fig. \ref{delta}. This Figure shows
that the location of the tricritical point does not depend very
much on the dimensionless parameter $\delta\Delta_0^6$. In fact
when it varies between zero and one, the temperature of the second
tricritical point varies of about 25\%, whereas the value of
$\delta\mu$ changes of about 1.6\%. The optimal value of $Q$ stays
practically constant in this range.

We can compare our results with the one obtained in \cite{matsuo}.
In this paper it has been found, by using the method of the
quasi-classical Green's function, that the second tricritical
point for the two antipodal plane waves is located at $T\approx
0.075\, T_c\approx 0.043\,\Delta_0$. By looking at  Figure
\ref{delta}, this suggests that the actual value of $\delta$ at
small temperature is rather small, but this conclusion should be
confirmed by an explicit evaluation of this parameter. Also it
should be noticed that the analysis made here using the
Ginzburg-Landau expansion along a second order line puts this
result in a more firm basis, and clarifies on a general ground the
existence of a second order transition point.

 \section{Conclusions}

 In this paper we have discussed the phase structure of the LOFF
 phase in the configuration of two antipodal plane waves at low
 temperature. The interest in this study is because it is
 known that at $T=0$ the system undergoes a second order phase
 transition at a critical value of the chemical potential
 separation $\delta\mu$ of the two pairing species of fermions. On
 the other hand the transition is first order at the tricritical point $P_{tric}$.
Here we clarify the reasons for this behavior as due to the
existence of a second tricritical point. The most important
 point of the paper is that, in the case of a Ginzburg-Landau expansion up to the eighth order
 in the gap, the phase space is essentially
 three-dimensional and that two tricritical lines exist. In fact
 we have studied in full generality the critical points in this
 space, under the assumption that the eighth term in the
 Ginzburg-Landau expansion is strictly positive. The existence of
 the second tricritical point is then a simple consequence of the
 general structure of the phase space when the eighth order term
 becomes important, and it comes about since starting from $T=0$ and
 increasing the temperature along the second order transition line, one follows a path
 in the physical phase space which crosses the second tricritical line.
 The study made in this paper,
 although
 motivated by a particular situation (the two antipodal plane wave of the LOFF phase),
 is in fact much more general and it is conceivable that it can be
 applied to other interesting physical situations.

\newpage

\end{document}